\documentclass{IEEEtran}
\usepackage{amsmath}
\usepackage{amssymb}
\usepackage{latexsym}
\usepackage{multirow}
\usepackage{epsfig}
\usepackage{graphics}
\usepackage{mathrsfs}
\usepackage{algorithmic}
\usepackage{algorithm}
\usepackage{graphicx}
\usepackage{subcaption}
\usepackage{epstopdf}
\usepackage{color}

\newcommand{\beq}{\begin{equation}}
\newcommand{\enq}{\end{equation}}
\newcommand{\ben}{\begin{eqnarray}}
\newcommand{\enn}{\end{eqnarray}}
\newcommand{\bea}{\begin{array}}
\newcommand{\ena}{\end{array}}
\newcommand{\bei}{\begin{itemize}}
\newcommand{\eni}{\end{itemize}}

\begin{document}
\begin{titlepage}
\title{Wearable Communications in 5G: Challenges and Enabling Technologies}
\author{Haijian Sun, Zekun Zhang, Rose Qingyang Hu, and Yi Qian

\thanks{Haijian Sun, Zekun Zhang, and Rose Qingyang Hu are with the Department of Electrical and Computer Engineering,
Utah State University, USA (e-mail: rosehu@ieee.org). Yi Qian is with the Department of Electrical and Computer Engineering, University of Nebraska-Lincoln, USA (e-mail: yqian@ieee.org). }
}
\end{titlepage}
\maketitle

\begin{abstract}
As wearable devices become more ingrained in our daily lives, traditional communication networks primarily designed for human being-oriented applications are facing tremendous challenges. The upcoming 5G wireless system aims to support unprecedented high capacity, low latency, and massive connectivity. In this article, we evaluate key challenges in wearable communications.  A cloud/edge communication architecture that integrates the cloud radio access network, software defined network,  device to device communications, and cloud/edge technologies  is  presented.  Computation offloading enabled by this multi-layer communications architecture can offload computation-excessive and latency-stringent applications to nearby devices through device to device communications  or to nearby edge nodes through cellular or other wireless technologies. Critical issues faced by wearable communications such as short battery life, limited computing capability, and stringent latency can be greatly alleviated by this cloud/edge architecture.  Together with the presented architecture, current transmission and networking technologies, including non-orthogonal multiple access, mobile edge computing, and energy harvesting, can greatly enhance the performance of wearable communication in terms of spectral efficiency, energy efficiency, latency, and connectivity.

\end{abstract}


\IEEEpeerreviewmaketitle

\section{Introduction}

Recent years have witnessed the unprecedented growth of wearable devices owing to the swift advances in chip design, computing, sensing and communications technologies. While wearable devices are not new, the past few years have seen a surge in their large-scale use and popularity.  It is estimated that close to 929 million devices will be available globally by 2021, a massive increase from 325 million in 2016 \cite{cisco}. A wearable device or simply a wearable refers to the device that can be worn on the body. This rapid rise in popularity was spurred, in part, by technological innovation. Emerging system on chip (SoC) and system in package (SiP) have  scaled down the printed circuit board (PCB) size, decreased power consumption, and most importantly,  have made it possible to design wearables in a variety of desired shapes.
\begin{figure}[h]
  \centering
  \includegraphics[width=3in]{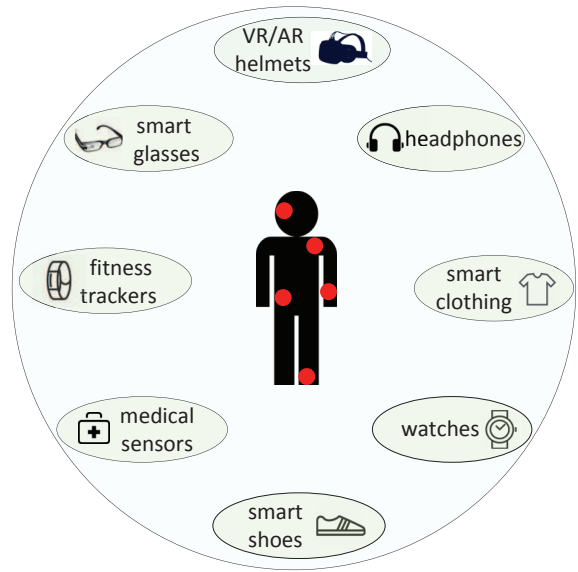}\\
  \caption{Wearable devices may have varying forms, from small medical sensors to entertainment helmets.}\label{1}
\end{figure}

Wearables provide easier access to information and convenience for their users. They have varying form factors, from low-end  health and fitness trackers to high-end virtual reality (VR) devices, augmented reality (AR) helmets, and smart watches. These devices can collect data on heart rates, steps, locations, surrounding buildings, sleeping cycles,  and even brain waves. Yet computing limitations continue to hinder wearables ability to process data locally.  As a result, most send their collected data to other powerful devices or to the clouds.  This necessary communication plays a key role in wearable devices and is the main emphasis in this paper. Different applications provided by different wearables may have varying communication requirements. For examples, while medical sensors have stringent requirements on latency and reliability, they have a low data rate need. On the other hand, AR/VR devices need both high throughput and low latency for a better user experience.

The upcoming 5G aims to support diverse communication requirements while serving as a unified platform for various services and applications. Before   5G finally takes over,  existing technologies  such as LTE-Advanced and wireless local area network (WLAN) are gradually evolving to fit new needs.  Furthermore, communication requirements of wearable devices can be fulfilled in part by existing technologies. For example, multi-user multiple-input-multiple-output (MU-MIMO), together with beamforming in 802.11ac, can achieve a throughput of over 1 Gbps \cite{WLAN}. For wearables requiring  high data rates, the future evolution of WLAN (IEEE 802.11 family in particular) can act as an alternative solution.  Perhaps most importantly, the hardware cost and power consumption in specifications like WLAN and Bluetooth are more suitable for wearable devices.
This article focuses on challenges as well as enabling technologies in wearable communications. The main contributions are as follows:

\begin{enumerate}
\item We evaluate design challenges and requirements for wearable communications and present a communication architecture that reflects recent industrial/academia research directions.
\item A list of detailed techniques are selected that can help alleviate the challenges. We emphasized on MAC/PHY design and present our research results accordingly.
\end{enumerate}

\section{Challenges for Wearable Communications}

\subsection{Power Constraints}
The relatively small size of most wearable devices poses a challenge when it comes to fitting a conventional battery inside. For most wearables, battery power tends to be proportional to the device size.  Hence today's consumer electronics design must reserve a vast amount of space for batteries alone. Some wearable devices also mount power consuming components like network chips, GPS, and continuous monitor sensors. For healthcare sensors, it is also important to keep the device unobtrusive to patients, especially for the implantable ones. The typical battery life should last at least several years. To make matters worse, wearable devices become fashion icons; hence, bulky and/or heavy design will inevitably flop in the market.

\subsection{Variations on Communication Requirements}
Wearable communication requirements vary depending on different use cases, in terms of differing data rate, latency, and reliability. On  the one hand, traffic growth has historically  been a key driving forces for new generation wireless systems. Since 2014,  VR/AR technologies have turned into a clear reality and  gigabit/s throughput has practically become a demanding feature in wearable device consumer markets. The  roadmaps of LTE and 5G have both proposed a gigabit/s experience in the near future. However, wearable devices may not be able to take full advantage of LTE and 5G, due to their potential cost and hardware complexity. On the other hand, wearable devices have succeeded in becoming more and more involved in everyday activities requiring voice, image, and video inputs. Human beings are generally sensitive to an approximate 100 $\rm{ms}$ audible delay and can catch visual delays of less than 10 $\rm{ms}$. Furthermore, cell phones and tablets now  use primarily touch interaction, a ``tactile interaction'' that requires a more rigorous  delay control, such as 1 $\rm{ms}$.

\subsection{Dense Deployment of Wearable Devices}
Wearables can help users see, hear, sense, and even feel the world, making it very common for one user to require multiple devices, such as a fitness tracker to maintain a healthy lifestyle, a VR/AR helmet for gaming and exploring a richer experience, and a smart glasses for virtual assistance and navigation. Such a usage scenario may not cause problems  in rural areas. Yet in areas with high population density, capacity and connectivity issues can become exaggerated and turn into  performance barriers \cite{dense}.

While most wearable communications occur locally by using WLAN, the contention based WLAN MAC protocol could limit the number of devices that can be supported. For example, even in ideal conditions, a typical Wi-Fi router can support a maximum of 30 to 50 connected devices.  With hundreds of people in a large conference room, the number of wearable devices could reach thousands. Not even a large deployment of hotspots could solve that communication problem, due to severe interference.

\subsection{Health Concerns}
A major concern  regarding wearable communications  is human biological safety under radio frequency (RF) exposure. The human body absorbs electromagnetic radiation,  which causes thermal or non-thermal heat in the affected tissues.  Guidelines on RF exposure normally apply specific absorption rate (SAR) as the metric for frequencies below 6 GHz. For mmWave, since the absorption is low and the primary energy remain in the surface layer of the skin, power density (PD) instead of SAR is more suitable for evaluating the health effects. However, PD cannot evaluate the effect of certain transmission characteristics such as reflection well. Therefore, temperature elevation of a direct contact area is proposed as the appropriate metric for mmWave RF exposure in \cite{health}. Besides, some tissues like eyes are especially vulnerable to mmWave radiation-induced heating and requires more attention.
It is necessary to continually update regulations based on new materials, frequencies, device types, and transmitted powers. In addition, manufacturers must be educated with the newest research/regulations to better mitigate consumer concerns and promote this new technology.

 \subsection{Security}
Due to the computing and power limitations of wearable devices, collected  data may need to be shared with other devices, edge nodes, or the cloud for further processing. The ever-growing desire to improve health and lifestyle remarkably promotes information sharing, such shared data will inevitably contain sensitive and private information, such as location, heart rate, emotion, and disease history. Thus, any leak of information could cause serious problems for individuals. The challenges here are multifold,  including how to protect data so that unauthorized people will not have access, how to ensure data is securely shared between  the device and the  cloud, and how to make sure data is securely stored in the cloud.

\section{Enabling Architecture for Wearable Communications}

\begin{figure*}[ht]
  \centering
  \includegraphics[width=0.8\textwidth]{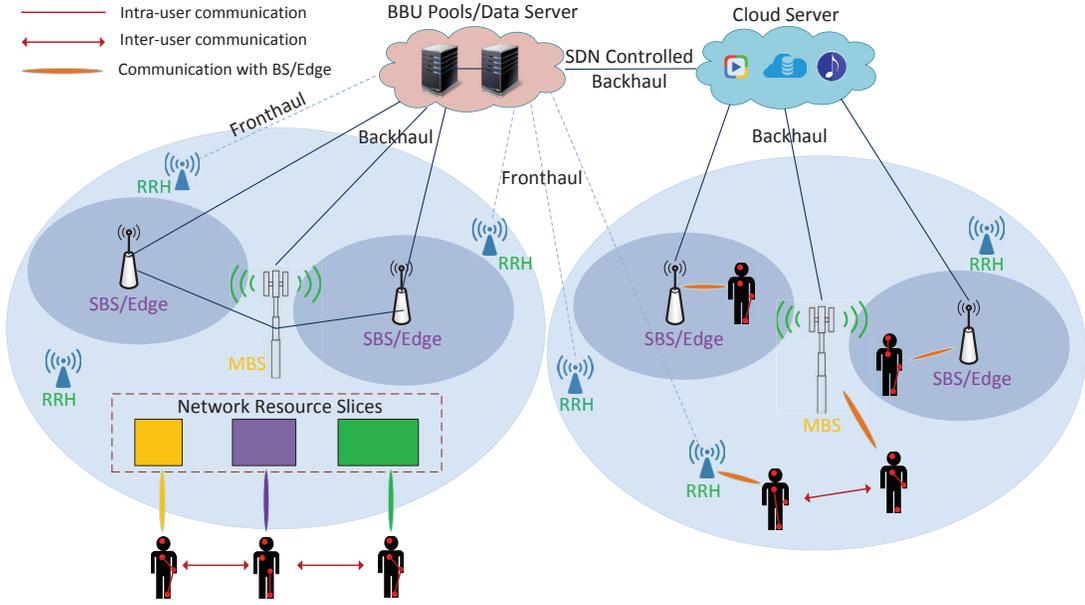}\\
  \caption {An illustration of the wearable communication system architecture.}\label{2}
\end{figure*}

This article presents a wearable communication architecture that combines heterogeneous cloud radio access networks (H-CRAN) \cite{HCRAN}, cloud/fog computing, and software defined networks (SDN),  as shown in Fig. \ref{2}. The H-CRAN architecture leverages macro base station (MBS), small base station (SBS), and remote radio header (RRH) to facilitate various user connections and performance needs. High power MBSs  provide blanket coverage and seamless mobility, while low power SBSs and RRHs enable local coverage and fulfill high capacity requirements \cite{hu_energy}. Wearables can utilize both MBSs and SBSs for data offloading, thereby saving energy and achieving faster computation speeds. Furthermore, MBSs and SBSs can connect to baseband unit (BBU) pools or to cloud servers directly, using backhaul connections. These BBU pools can help achieve globally optimized mobile association, interference management, and cooperation. When MBSs/SBSs are directly connected to cloud servers, extremely computation-intensive but less delay-stringent tasks  can be offloaded to cloud servers for more powerful processing. To further leverage cloud RAN benefits, RRHs can be set up very close to end wearable devices. RRHs can be designed to mainly possess radio front functionalities while BBU pools handle the majority PHY/MAC layer processing.  Low power RRHs  can be largely deployed to provide various communication needs, such as low latency, low transmission power,  and high capacity. In addition, RRHs can integrate several transmission technologies, including Bluetooth, WLAN, and visible light communication (VLC), to  help provide backward compatibility and enrich use cases.

BBU pools are connected by data servers, in which user-specific data, such as preferences, locations, activities predictions, and quality of service (QoS) requirements, are stored. Data can be generated by cloud servers with SDN-controlled backhaul, wherein SDN controllers play an important role. To be specific, SDN controllers are aware of network condition and user demands, send instructions to BBU pools to guide network traffic forwarding. Furthermore, local SDN controllers in each sub-network can abstract physical devices to virtual ones. For example, depending on the application, network resources can be sliced to form virtual access points (APs) to fit different needs \cite{SDN}. Such network slicing could dynamically improve system performance.

Another important component is mobile edge computing (MEC). MEC is very important to wearable communications because it moves clouds locally to reduce transmission latency, backhaul loading, and the central node workload. One observation in wearable devices is that a significant portion of communications take place between various devices belonging to the same person. Therefore,  device-to-device (D2D) communications underlaying cellular network can transmit and process data  locally,  thereby reducing both latency and energy consumption. This cloud/edge  architecture with D2D could greatly facilitate wearable communications.  Communications between wearables and edge nodes can use different technologies, either on licensed bands or unlicensed bands.
   \begin{table}[hbt]
   \centerline{\psfig{file=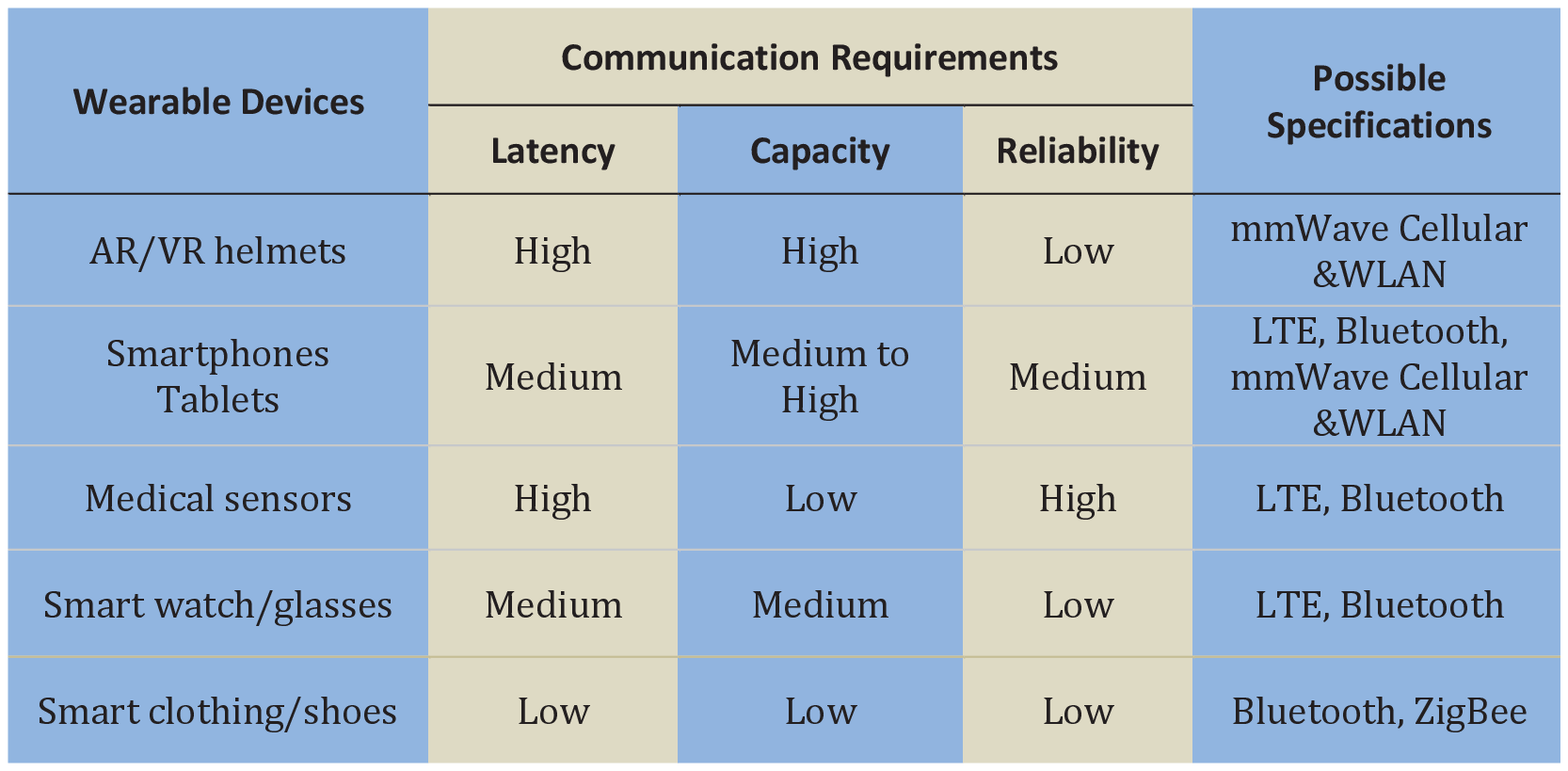,width=3.6in}}
   \renewcommand{\baselinestretch}{1}\normalsize
   \caption{Wearable communication requirements and possible solutions}
   \label{table1}
   \end{table}

\begin{enumerate}
\item Licensed Wearable Communications:
Commercial cellular communications generally fall into this category. Within licensed communications,  wearables can communicate with either edge nodes or BSs directly. Although QoS and mobility can be well supported through cellular networks, some potential drawbacks may exist. First, exclusive usage of  spectrum elevates the expedient cost on carriers and ultimately increases customer cost. In addition, licensed communication chips tend to be more complex and expensive. Further, licensed communication often consumes more power.

\item Unlicensed Wearable Communications:
2.4 GHz and 5 GHz WLAN, Bluetooth, and IEEE 802.15.4 are the most prevalent Sub-6 GHz unlicensed technologies used in today's wearable devices, with IEEE 802.11 ad (WiGig, 60 GHz WLAN) and visual light communications (VLC) still under research or under deployment. These specifications allow close proximity direct communications between two or more devices. Due in part to these factors, unlicensed communications enable cheaper and less complex devices, as well as a longer battery life, all of which are desirable for wearables. However, the maximum transmitted power constraint confines communications to a limited range at the unlicensed band. Furthermore, contention based access schemes of some techniques may impose certain limitations on the number of devices supported.
\end{enumerate}
Table. \ref{table1} lists all the communication specifications and their potential use cases in wearable communications.
\section{Enabling Transmission/Networking Technologies}

\subsection{Antenna Design}

The form factor and power constraints on wearable devices impose additional requirements on  antenna design. This is especially true for those devices operating on multiple modes, with transceivers designed to work in more than one protocol stack. As previously mentioned, differing  wireless technologies may use  differing spectrum. Traditional cellular, Bluetooth, tri-band Wi-Fi (2.4 GHz, 5 GHz, and 60 GHz) and mmWave cellular are expected in the high-end wearable devices by the consumer market. Yet legacy antenna design requires the antenna size to be less than half of the wavelength, which can efficiently capture the radiated signal. For cellular systems, the frequency ranges from 800 MHz (GSM, low band 3G, and 4G) to around 2.5 GHz (high band 4G, Bluetooth, and 2.4 GHz Wi-Fi),  antenna size varies from 18 cm to 5 cm. Today's innovative antenna design incorporates both engineering and industrial design aspects, that take advantage of the entire structure of a device. For smaller devices, patch antenna can be directly printed on the circuit board with a higher dielectric constant, thus reducing the required size at the cost of gain  loss. Besides, \cite{Antenna} showed an innovative tri-band antenna design that can work in small-size wearables.

 The mmWave frequency band not only leads to higher path loss, it is more susceptible to blockages, potential water vapor, and oxygen molecule absorption.  However, a smaller wavelength at the mmWave band can benefit the antenna array design within a compact area. Hence, for mmWave communications, an antenna array with multiple antenna elements and directional beamforming could compensate for the downside of channel characteristics. In fact, as RF units drain a significant amount of battery energy when compared with other antenna components, the most advanced designs use fewer RF units, while still achieving a promising performance.  The idea is to group several antenna elements into a single RF unit, thereby leading to a hybrid analog/digital antenna structure. Due to the sparse nature of multipath in mmWave signal, an improvement from pure digital beamforming is limited. And the relative simplicity of analog  beamforming further motivates the hybrid analog and digital antenna structure. A prototype design made by Samsung Electronics has 32 antenna elements but only 4 RF units in a 6 cm $\times$ 3 cm area \cite{samsung}. Such architecture can easily be applied to wearable devices such as AR helmets.

  With regards to the base station (BS), a large-scale antenna system (typically in the order of hundreds) can be mounted to form massive MIMO systems capable of serving more users with the same time/frequency resources and provide a higher energy and spectral efficiency. Massive MIMO takes greater advantage of spatial diversity and/or multiplexing by supporting massive connectivities in a scalable way. The gain, however, comes primarily from an accurate knowledge of the channel state information (CSI)  used for signal detection and precoding. How to obtain CSI using moderate to low overhead with pilot contamination in a massive MIMO system is an active yet challenging research. For wearable devices, due to their close proximity to the human body, antenna design must further consider the SAR. This regulation could impact the power level and antenna beam design.

 \subsection{PHY and MAC Technologies}

   The massive connectivity and high data rate requirements of wearable devices can be fulfilled, in part, by new radio access technologies (RATs) and MAC technologies. Emerging RATs, such as non-orthogonal multiple access (NOMA), benefit the system with both spectral efficiency and connectivity \cite{vtc_d2d}. NOMA allows the same radio resources to be used by more than one wearable device at the same time, a contrast to current orthogonal  multiple access (OMA) technologies, such as orthogonal frequency-division multiple access (OFDMA) in 4G. The non-orthogonality can  occur either in the power-domain (PD-NOMA) or the code domain (CD-NOMA). CD-NOMA utilizes different codes within the same resource to achieve multiplexing gain, while PD-NOMA assigns users with distinct power levels to maximize the performance.
    \begin{figure}[h]
  \centering
  \includegraphics[width=3.6in]{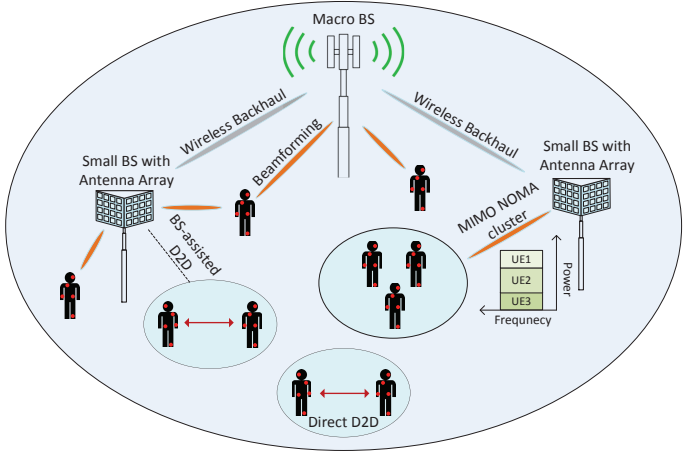}\\
  \caption{A wearable communication  system with MIMO, NOMA, and  D2D PHY/MAC schemes}\label{3}
\end{figure}
    In this article, we focus on PD-NOMA, denoted as NOMA for brevity. On the transmitter side, NOMA allocates more power to users with poor channel conditions, creating a power disparity which not only facilitates decoding, but also promotes system throughput and fairness. NOMA has the potential to achieve even higher spectral efficiency and connectivity at the cost of a more complex receiver structure, a problem which can be addressed by more advanced signal processing schemes and hardware design. To be specific, SIC is used to retrieve user messages by decoding the strongest signal first. SIC then subtracts the decoded  signals  and continue to decode the next strongest signal. This process stops once the intended received signal is decoded. Furthermore, D2D communication is considered a promising technology  in 5G systems \cite{lili_d2d}. The D2D underlying cellular network allows direct communications between closely located users. Both D2D assisted cellular mode communications and direct D2D communications can utilize close proximity and frequency reuse gains so that higher energy and spectral efficiency can be achieved. A wearable communication system using  NOMA,  MIMO, and D2D PHY/MAC schemes  is shown in Fig. \ref{3}.

\begin{figure*}
  \centering
  \includegraphics[width=\textwidth]{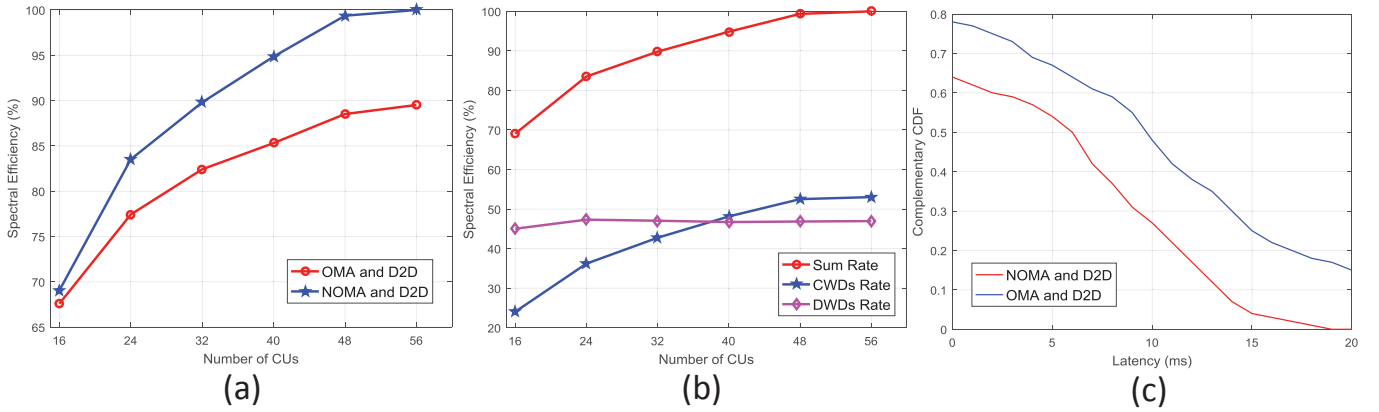}\\
  \caption{Performance evaluation of a downlink system with MU-MIMO, NOMA and D2D. $R = 1$ $\rm{km}$, $M=4$, $K=2$, $D=2$, $R_d = 10$ $\rm{m}$. Transmit powers of the edge node and DWDs are 10 and 2 Watts, respectively. (a), Sum rate of proposed NOMA+D2D with OMA+D2D. (b) The performance of CWDs and DWDs in NOMA+D2D scheme \cite{vtc_d2d}. (c) CCDF performance w.r.t the latency. }\label{4}
\end{figure*}

Fig. \ref{4}  presents preliminary simulation results when  multiuser-MIMO (MU-MIMO), NOMA, and D2D are applied to  wearable communications.  Specifically, consider a downlink wireless system with one edge node base station (BS) located in the center of a circle with a radius of $R$ km. The BS has $M$ antennas, whereas the cellular mode wearable devices  (CWDs) and $D$ pairs of D2D mode wearable devices (DWDs) have only one antenna each. Notice that a wearable can operate in both modes, depends on the connection requirement. CWD refers to the wearable that connect to cellular base stations for guaranteed service, while DWD is the mode for local connections.  CWDs and DWDs are randomly deployed. The distance between each D2D transmitter and receiver is constant and denoted as $R_d$. The channel gain between the BS, CWDs, and DWDs consists of large-scale path loss and small-scale Rayleigh fading. To better utilize MIMO and NOMA, the  edge node generates $M$ beams and each beam supports $K$ CWDs through NOMA. Thus in total $M\times K$ users can be supported on each radio resource unit.
Precoding scheme and power allocation need to be optimized for maximizing the sum spectral efficiency of CWDs and DWDs. While the joint optimization is difficult to achieve, we solve the problem in a heuristic way,  wherein zero-forcing (ZF) precoding is determined first, followed by the NOMA power allocation is decided by applying Karush-Kuhn-Tucker (KKT) conditions in convex optimization. As a comparison, results from orthogonal multiple access (OMA) are also presented, with only one CWD supported  in each beam. All the results are expressed as the percentage for better illustration. Clearly, the proposed scheme reveals a better performance in terms of overall spectral efficiency, connectivity, and latency. And as the number of CWDs increases,  the system can further benefit from multiuser diversity gain.

The second part is an advanced MAC protocol which coordinates transmission/processing within a device or between devices. Traditional methods dispatch transmissions according to varying protocol stacks within a device so that data going  through BLE will not be sent to Wi-Fi. However, coordination can be made with a central unit (a dedicated low power always-on component) inside which takes a step further to send data to appropriate protocols, based on the availability, surrounding interference level, and demand for quality of experience (QoE). For example, emergency healthcare information can be dispatched to a cellular unit, resulting a fast response directly over the Internet. Voice calls can go through Wi-Fi if the cellular unit is unavailable. Besides, MAC plays a more important role in transmissions involving multiple devices.  Smart transmission classifies data in terms of QoS requirements, which can help save battery life by forcing the RF unit to enter a sleep mode that only activates to deliver critical information requiring low latency transmission. Further, since antennas can form narrow beams at the  mmWave band,  devices can support multiple transmissions simultaneously with limited co-channel interference \cite{LLWei}. MAC protocol needs to consider initial access aligning antenna to the high gain direction, thereby enabling beam-tracking in motions for seamless experience and an advanced sensing algorithm which senses the channel in a specified direction, rather than simply isotropically.

    \subsection{Cloud/Edge Computing}

\begin{figure}
  \centering
  \includegraphics[width=3.5in]{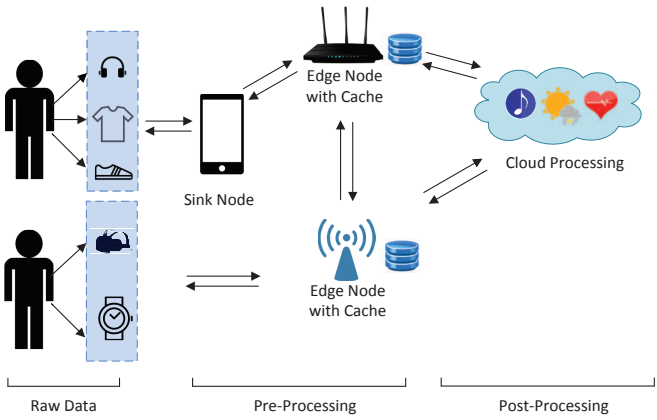}\\
  \caption{Edge communication overview}\label{5}
\end{figure}

Cloud computing has brought significant changes to the Internet in the past few decades. Its centralized nature helps lower the expenditure cost while speeding up the deployment process.  However, cloud computing alone cannot fulfill the demands of wearable communication.  Cloud data centers are often located in remote regions, which may cause a long end-to-end latency, thereby impacting delay-sensitive applications. Since data are sent to the cloud for  processing, concerns such as  security and privacy possibly may arise as well. Yet current research now shifted to a  combination of cloud and edge computing structure. Specifically, devices or nodes with storage, computing, and caching capabilities can be deployed in close proximity with wearable devices and act as middleware between cloud and local networks. These devices can be routers, small base stations, and even high-end wearable devices. In addition, advanced caching algorithms can offload popular contents from cloud to edge nodes either in real-time or offline. An illustration of the cloud/edge architecture is shown in Fig. \ref{5}. To better take advantage of varying spectrum, connections between edge nodes could utilize mmWave band which provides sufficient bandwidth for higher throughput \cite{dense_wireless}, while the connection between devices belonging to the same person could use 2.4 GHz BLE and WLAN, or 5 GHz WLAN. By properly assigning spectrum, interference in the dense wearable networks can be reduced.

The advantages of this paradigm are multi-fold. Firstly, by providing certain computing capabilities via edge nodes or wearable devices, the transmission load  on the backhaul can be greatly reduced. This benefit is prominent for applications such as online gaming, where 60 or even 120 frames need to be rendered per second. An alternative solution dictates that, servers only send parameters such as character's  position, time-stamp, and property changes (few plain data) and allow the edge nodes to calculate and  render visual images. Secondly, with the help of the large number of edge nodes deployed in 5G  and big data analysis of user preferences, popular contents can be pre-fetched into connected edge devices, which are only one hop away from users. Thirdly, this scheme is more robust in terms of always-on connectivity, as well as privacy and security control. Lastly, cloud/edge computing enables a much more scalable architecture.

    \subsection{Energy Harvesting}

The advancement of battery technologies lags behind its silicon counterparts. Nowadays, widely-used consumer device batteries are based on Lithium-ion.  Researchers are working on improving battery energy density, finding new materials, and reducing charging time to deliver a better user experience.  Meanwhile, energy efficiency has become a major concern in the network design. This problem can be alleviated by developing advanced  energy harvesting techniques, which enable devices to harvest energy from the surrounding environment for both immediate and/or future usage by storing harvested energy in the battery unit. Such energy can  come from solar power, ambient  motion, the human body, background electromagnetic waves, etc. Solar power, for example, can be used to run outdoor wearables, such as edge nodes, watches, and smart clothes.  Recent progress on solar cell materials like perovskites, make the solar power harvest more flexible to integrate, and more efficient, not to mention cheaper. Ambient motion takes the advantage of mechanical movements by transferring them to electrical form. In general, direct force and inertial force on a proof mass are two main energy sources. Their principles, however, are similar. Since the generated energy of this technique is relatively low (a few microwatts, depending on specific activities), it is more promising for applications like foot-wear equipment and watches \cite{EH1}. Furthermore, wearable devices can extract energy from the human body by capturing temperature differences between the body and the outside environment with a thermoelectric module. Even though the efficiency is limited, with only a few Celsius difference on average, its value has been proved by various commercialized products. In addition, studies have also reported the  energy harvest from human body fluids. Lastly, energy harvest from electromagnetic waves is attracting more attention recently.

\subsection{Advanced Security Solutions}

\begin{figure}[!t]
  \centering
  \includegraphics[width=3.5in]{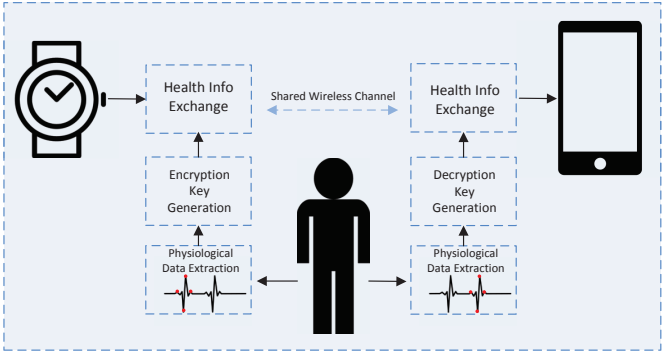}\\
  \caption{An example of intra-wearable security solution using heart rate pattern}\label{6}
\end{figure}

Concerns with data security and privacy have increased, especially as users share more and more private data, including photos, locations, and activities. However, physical data collected from wearable devices such as medical conditions are sensitive and require extra protection. Typically, data goes through different phases, namely, data collection, transmission, and sharing.  In data collection, biometric access is already widely used in high-end devices. Iris, face, and fingerprint recognition utilize specific user patterns to secure device access.  Data can be further secured with schemes  such as  public key encryption.  For some wearable applications, data needs to be shared to remote servers for analysis or diagnosis.  Encryption requires collaboration  with network architecture and transmission protocols. To be specific, we consider intra-wearable and inter-wearable communication scenarios.    Intra-wearable communications  occur  between multiple devices carried by the same person while inter-wearable communications  occur  between multiple devices carried by different people. In the former case,  biometric information such as inter-pulse interval (IPI) can be easily detected by multiple devices and can then be extracted for encryption and decryption key generation \cite{Encryption}. This extra protection at the protocol level makes wearable communications more secure. A brief illustration is in Fig. \ref{6}. Inter-wearable communications,  on the other hand, can leverage edge nodes and cloud servers.  Specifically, public key cryptography can be made between wearables and servers for scalable considerations. For resource-constrained wearables, the impact of computational workload and power consumption on security should also be taken into consideration.

\section{Conclusions}
Recent explosive growth of wearable devices has spurred ever-increasing research interests in various fields, including communications. Yet such growth also presents paramount challenges in the same fields. In order to tackle these challenges, communications architecture and communication technologies are contemplating revolutionary changes.  The multiple-layer communication architecture presented in this article combines D2D, C-RAN, and cloud/edge technologies into one to address stringent latency/power/computation constraints in wearable communications.  Enabled by this multi-layer communications architecture, computation offloading to nearby devices, through D2D or to nearby edge nodes through cellular/other wireless technologies  has been deemed  one of the key techniques used to address fundamental wearable issues, such as limited battery, limited computing capability, critical latency on performance.   Transmission technologies such as massive MIMO and NOMA applied wearable communications,  can further significantly improve wearable communication spectral efficiency, power efficiency, and connectivity.

\section{Author Information}
\emph{\textbf{Haijian Sun}} (h.j.sun@ieee.org) received the B.S. and M.S. degrees from Xidian University, Xi'an, China, in 2011 and 2014, respectively. He is currently pursuing the Ph.D. degree with the Department of Electrical and Computer Engineering, Utah State University, Logan, UT, USA. His research interests include MIMO, nonorthogonal multiple access, SWIPT, wearable communications, and 5G PHY.

\emph{\textbf{Zekun Zhang}} (zekun.zhang.z@ieee.org) received the B.S. degree in electronics and information engineering from Beihang University, Beijing, China, in 2012. He is currently pursuing the Ph.D. degree with the Department of Electrical and Computer Engineering, Utah State University. His current research interests include device to device communication, wireless heterogeneous networks, and nonorthogonal multiple access systems.

\emph{\textbf{Rose Qingyang Hu}} (rosehu@ieee.org) is a Professor of Electrical and Computer Engineering Department at Utah State University. She received her B.S. degree from University of Science and Technology of China, her M.S. degree from New York University, and her Ph.D. degree from the University of Kansas. She has more than 10 years of R\&D experience with Nortel, Blackberry and Intel as a technical manager, a senior wireless system architect, and a senior research scientist, actively participating in industrial 3G/4G technology development, standardization, system level simulation and performance evaluation. Her current research interests include next-generation wireless communications, wireless system design and optimization, green radios, Internet of Things, Cloud computing/fog computing, multimedia QoS/QoE, wireless system modeling and performance analysis. She has published over 180 papers in top IEEE journals and conferences and holds numerous patents in her research areas. Prof. Hu is an IEEE Communications Society Distinguished Lecturer Class 2015-2018 and the recipient of Best Paper Awards from IEEE Globecom 2012, IEEE ICC 2015, IEEE VTC Spring 2016, and IEEE ICC 2016.

\emph{\textbf{Yi Qian}} (yqian@ieee.org) received a Ph.D. degree in electrical engineering from Clemson University. He is currently a professor in the Department of Electrical and Computer Engineering, University of Nebraska-Lincoln (UNL). Prior to joining UNL, he worked in the telecommunications industry, academia, and the government. Some of his previous professional positions include serving as a senior member of scientific staff and a technical advisor at Nortel Networks, a senior systems engineer and a technical advisor at several start-up companies, an assistant professor at University of Puerto Rico at Mayaguez, and a senior researcher at National Institute of Standards and Technology. His research interests include information assurance and network security, network design, network modeling, simulation and performance analysis for next generation wireless networks, wireless ad-hoc and sensor networks, vehicular networks, smart grid communication networks, broadband satellite networks, optical networks, high-speed networks and the Internet. Prof. Yi Qian is a member of ACM and a senior member of IEEE. He was the Chair of IEEE Communications Society Technical Committee for Communications and Information Security from January 1, 2014 to December 31, 2015. He is a Distinguished Lecturer for IEEE Vehicular Technology Society and IEEE Communications Society. He is serving on the editorial boards for several international journals and magazines, including serving as the Associate Editor-in-Chief for IEEE Wireless Communications Magazine. He is the Technical Program Chair for IEEE International Conference on Communications (ICC) 2018.

\end{document}